\documentclass[12pt]{article}

\begin{document}

\title{Scattering by $\mathcal{P}\mathcal{T}$-symmetric Non-local Potentials 
}
\author{Francesco Cannata \\
Istituto Nazionale di Fisica Nucleare, Sezione di Bologna\\
and Dipartimento di Fisica dell' Universit\`a,\\
Via Irnerio 46, I 40126 Bologna, Italy \\
\and Alberto Ventura \\
Ente Nuove Tecnologie, Energia e Ambiente, Bologna\\
and Istituto Nazionale di Fisica Nucleare, Sezione di Bologna, Italy}
\maketitle

\begin{abstract}
A general formalism is worked out for the description of one-dimensional
scattering by non-local separable potentials and constraints on transmission
and reflection coefficients are derived in the cases of $\mathcal{P}$, $%
\mathcal{T}$ or $\mathcal{PT}$ invariance of the Hamiltonian. The case of a
solvable Yamaguchi potential is discussed in detail.

PACS numbers: 03.65.Ca, 03.65.Nk \newline
\emph{Keywords:} non-hermitian quantum mechanics, non-relativistic scattering
\end{abstract}

\section{Introduction}

\label{intro}

Non local potentials (see, \textit{e.g., }Ref.\cite{Ao83}) play an important
role in several applications of quantum scattering theory. In nuclear
physics, for instance, \ they naturally arise from convolution of an
effective nucleon-nucleon interaction with the density of a target nucleus.
In the present note, focused on one-dimensional scattering, we study the
behaviour of reflection and transmission coefficients of a non-local
solvable potential with separable kernel in connection with the
characteristics of the kernel itself, by considering, in particular, the
cases where the kernel is real symmetric, hermitian or $\mathcal{PT}$%
-symmetric. \ 

\section{Formalism}

\label{nonloc}

The formalism adopted in the present work is consistent with recent general
references on one-dimensional scattering by complex potentials, such as Ref.%
\cite{Mu04}, \ and $\mathcal{PT}$-symmetric potentials, such as Ref.\cite%
{CDV06}.

Let us introduce the one-dimensional Schr\"{o}dinger equation for a
monochromatic wave of energy $E=k^{2}$ scattered by a non-local potential
with kernel $K$, written in units $\hbar =2m=1$ 
\begin{equation}
-\frac{d^{2}}{dx^{2}}\Psi (x)+\lambda \int K(x,y)\Psi (y)dy=k^{2}\Psi (x)\,,
\label{Schrnl}
\end{equation}%
where the potential strength, $\lambda $, is a real number.

It is easy to check, by calculating scalar products, that the kernel of a
hermitian non-local potential satisfies the condition 
\begin{equation}
K(x,y)=K^{\ast }(y,x)\,.  \label{hermnl}
\end{equation}

Parity \ ($\mathcal{P}$) invariance of the potential could be similarly
checked to imply 
\begin{equation}
K(x,y)=K(-x,-y)\;.  \label{Parinvnl}
\end{equation}

The condition of time reversal \ ($\mathcal{T}$) invariance of $K$ can be
written in the form 
\begin{equation}
K(x,y)=K^{\ast }(x,y)\;,  \label{Tinvnl}
\end{equation}%
while $\mathcal{PT}$ \ invariance corresponds to the condition

\begin{equation}
K(x,y)=K^{\ast }(-x,-y)\;,  \label{ptnl}
\end{equation}%
in agreement with formula (3) of Ref.\cite{Ru05} , which corrects a misprint
in the corresponding formula (113) of Ref.\cite{Mu04}.

In order to deal with a solvable $\mathcal{PT}$-symmetric potential, we
consider only separable kernels of the kind 
\begin{equation}
K(x,y)=g(x)e^{i\alpha x}h(y)e^{i\beta y}\,,  \label{K_pm}
\end{equation}
where $\alpha $ and $\beta $ are real numbers, and $g(x)$ and $h(y)$ are
real functions of their argument, suitably vanishing at $\pm \infty $.

For this kind of kernel, the hermiticity condition (\ref{hermnl}) implies $%
\alpha =-\beta $ and $g=h$. Parity invariance (\ref{Parinvnl}) requires $%
\alpha =\beta =0$ and $g\left( x\right) =g\left( -x\right) $, $h\left(
x\right) =h\left( -x\right) $. Time reversal invariance (\ref{Tinvnl})
requires $\alpha =\beta =0$ , but does not impose conditions on $g$ and $h$.

The various conditions that can be imposed on kernel (\ref{K_pm}) are
summarized in Table 1. 
\begin{table}[tbp]
\begin{center}
\begin{tabular}{|c|c|}
\hline
Reality & $\alpha =\beta =0$ \\ \hline
Symmetry under $x\leftrightarrow y$ & $\alpha =\beta $, $g=h$ \\ \hline
Hermiticity & $\alpha =-\beta $, $g=h$ \\ \hline
$\mathcal{P}$ Invariance & $\alpha =\beta =0$, $g\left( x\right) =g\left(
-x\right) $, $h\left( y\right) =h\left( -y\right) $ \\ \hline
$\mathcal{T}$ Invariance & $\alpha =\beta =0$ \\ \hline
$\mathcal{PT}$ Invariance & $g\left( x\right) =g\left( -x\right) $, $h\left(
y\right) =h\left( -y\right) $ \\ \hline
\end{tabular}%
\end{center}
\caption{Possible symmetries of the separable kernel (\protect\ref{K_pm})}
\end{table}
Finally, $\mathcal{PT}$ invariance (\ref{ptnl}) does not impose conditions
on $\alpha $ and $\beta $, but requires $g(x)=g(-x)\,,h(y)=h(-y)\,$. As an
important consequence, their Fourier transforms, $\tilde{g}(q)$ and $\tilde{h%
}(q^{\prime })$, are real even functions, too. 

In order to solve eq. (\ref{Schrnl}), we resort to the Green's function
method. As is known, the Green's function of the problem is a solution to
Eq. (\ref{Schrnl}) with the potential term replaced with a Dirac delta
function 
\begin{equation}
\frac{d^{2}}{dx^{2}}G_{\pm }(x,y)+(k^{2}\pm i\varepsilon )G_{\pm
}(x,y)=\delta (x-y)\,.  \label{Helmeq}
\end{equation}

Here, we introduce the infinitesimal positive number $\varepsilon$ in order
to shift upwards, or downwards in the complex momentum plane the
singularities of the Fourier transform of the Green's function, $%
G_{\pm}(q,q^{\prime})$, lying on the real axis.

In fact, after defining the Fourier transform, $\tilde{f}(q)$, of a generic
function $f(x)$ as 
\[
\tilde{f}(q)=\int_{-\infty}^{+\infty}f(x)e^{-iqx}dx\quad \leftrightarrow
\quad f(x)=\frac{1}{2\pi} \int_{-\infty}^{+\infty}\tilde{f}(q)e^{iqx}dq \,, 
\]
and expressing $G_{\pm}(x,y)$ and $\delta(x-y)$ in terms of their Fourier
transforms, we quickly solve eq. (\ref{Helmeq}) for $G_{\pm}$ 
\[
\tilde{G}_\pm(q,q^{\prime})=\frac{2\pi\delta(q+q^{\prime})}{-q^2+k^2\pm
i\varepsilon}, 
\]

Therefore, the Green's function in coordinate space is 
\begin{equation}
G_{\pm }(x,y)=-\frac 1{2\pi }\int_{-\infty }^{+\infty }\frac 1{q^2-k^2\mp
i\varepsilon }e^{iq(x-y)}dq=G_{\pm }(x-y)\,.  \label{Greenf}
\end{equation}
The integral (\ref{Greenf}) is easily computed by the method of residues. In
fact, the integrand in $G_{+}(x-y)$ has two first order poles at $%
q_1=k+i\varepsilon ^{\prime }$ and $q_2=-k-i\varepsilon ^{\prime }$, where $%
\varepsilon ^{\prime }=\varepsilon /(2k)$: the integral is thus computed
along a contour made of the real $q$ axis and of a half-circle of infinite
radius in the upper half-plane for $x-y>0$, on which the integrand vanishes,
thus enclosing the pole at $q=q_1$, and in the lower half-plane for $x-y<0$,
enclosing the pole at $q=q_2$, for the same reason. Notice that the $G_{+}$
contour integral is done in the counterclockwise direction for $x-y>0$,
while it is done in the clockwise direction for $x-y<0$, so that the latter
acquires a global sign opposite to the former. The result is 
\begin{equation}
G_{+}(x-y)=-\frac i{2k}\left[ e^{ik(x-y)}\theta (x-y)+e^{-ik(x-y)}\theta
(y-x)\right] \,=-\frac i{2k}e^{ik\left| x-y\right| },  \label{G_p}
\end{equation}
where $\theta (x)$ is the step function, equal to 1 for $x>0$ and 0
otherwise.

The second Green's function, $G_{-}(x-y)$, is the complex conjugate of $%
G_{+}(x-y)$ 
\begin{equation}
G_{-}(x-y)=\frac i{2k}\left[ e^{-ik(x-y)}\theta (x-y)+e^{ik(x-y)}\theta (y-x)%
\right] =\frac i{2k}\,e^{-ik\left| x-y\right| }.  \label{G_m}
\end{equation}

Now, we go back to eq. (\ref{Schrnl}) with kernel (\ref{K_pm}), call $%
\Psi_\pm(x)$ two linearly independent solutions, for a reason that will
become clear in the next few lines, and define the following integral
depending on $\Psi_\pm$ 
\[
I_\pm(\beta,k)=\int_{-\infty}^{+\infty}e^{i\beta y} h(y)\Psi_\pm(y)dy\,. 
\]
It is easy to show that $I_\pm(\beta,k)$ can be written as a convolution of
the Fourier transforms of $h(y)$ and $\Psi_{\pm}(y)$.

The general solution to eq. (\ref{Schrnl}) is thus implicitly written as 
\begin{equation}
\Psi_\pm (x)= c_\pm e^{ikx}+d_\pm e^{-ikx}+\lambda I_\pm (\beta,k)
\int_{-\infty}^{+\infty}G_\pm (x-y)g(y)e^{i\alpha y}dy\,.  \label{Psipm}
\end{equation}

Eq. (\ref{Psipm}) allows us to express $I_{\pm }(\beta ,k)$ in terms of the
constants $c_{\pm }$ and $d_{\pm }$ and of Fourier transforms of known
functions : in fact, by multiplying both sides by $h(x)e^{i\beta x}$ and
integrating over $x$, we obtain 
\begin{equation}
I_{\pm }(\beta ,k)=c_{\pm }\tilde{h}(k+\beta )+d_{\pm }\tilde{h}(k-\beta
)+\lambda N_{\pm }(\alpha ,\beta ,k)I_{\pm }(\beta ,k)\,,  \label{Ipm1}
\end{equation}
where we have exploited the symmetry $\tilde{h}(-k-\beta )=\tilde{h}(k+\beta
)$ and $N_{\pm }$ is defined as 
\begin{eqnarray}
N_{\pm }(\alpha ,\beta ,k) &=&\int_{-\infty }^{+\infty }h(x)e^{i\beta
x}G_{\pm }(x-y)g(y)e^{i\alpha y}dxdy \\
&=&\mp \frac i{2k}\int_{-\infty }^{+\infty }h(x)e^{i\beta x}e^{\pm ik\left|
x-y\right| }g(y)e^{i\alpha y}dxdy,  \nonumber  \label{Npm}
\end{eqnarray}
so that 
\begin{equation}
I_{\pm }(\beta ,k)=\frac{c_{\pm }\tilde{h}(k+\beta )+d_{\pm }\tilde{h}%
(k-\beta )}{1-\lambda N_{\pm }(\alpha ,\beta ,k)}=(c_{\pm }\tilde{h}(k+\beta
)+d_{\pm }\tilde{h}(k-\beta ))D_{\pm }\,,  \label{Ipm2}
\end{equation}
where 
\[
D_{\pm }(\alpha ,\beta ,k)\equiv \frac 1{1-\lambda N_{\pm }(\alpha ,\beta
,k)}\,. 
\]

Let us examine now the asymptotic behaviour of the two independent
solutions, starting from $\Psi_+(x)$ 
\begin{equation}
\Psi_+(x) = c_+ e^{ikx}+d_+ e^{-ikx}+\lambda
I_+(\beta,k)\int_{-\infty}^{+\infty}G_+ (x-y)g(y)e^{i\alpha y}dy \, .
\label{Psi1}
\end{equation}
The asymptotic behaviour of the integral on the r. h. s. of eq. (\ref{Psi1})
is promptly evaluated by observing that, according to eq. (\ref{G_p}), 
\[
\lim_{x\rightarrow \pm \infty}G_+ (x-y)=-\frac{i}{2k}e^{\pm ik(x-y)}\,, 
\]
so that 
\[
\lim_{x\rightarrow \pm \infty}\Psi_+ (x)=c_+ e^{ikx}+d_+e^{-ikx}-i\omega I_+
(\beta,k)\tilde{g}(k\mp \alpha)e^{\pm ikx}\,, 
\]
where we have put $\omega=\lambda/(2k)$.

Remembering the expression (\ref{Ipm2}) of $I_+$, we finally obtain 
\begin{eqnarray*}
\lim_{ x\rightarrow -\infty}\Psi_+ (x) &=& c_+ e^{ikx}+\left\{d_+ -i\omega%
\tilde{g}(k+\alpha) \left[c_+\tilde{h}(k+\beta)+d_+\tilde{h}(k-\beta)\right]%
D_+\right\} e^{-ikx}\,, \\
\lim_{ x\rightarrow +\infty}\Psi_+ (x) &=& \left\{ c_+ -i\omega\tilde{g}%
(k-\alpha) \left[c_+\tilde{h}(k+\beta)+d_+\tilde{h}(k-\beta)\right]%
D_+\right\}e^{ikx} +d_+ e^{-ikx}\,.
\end{eqnarray*}

The constants $c_{+}$ and $d_{+}$ are fixed by initial conditions: if we
impose that $\Psi _{+}(x)$ represents a wave travelling from left to right,%
\begin{eqnarray*}
\lim_{x\rightarrow -\infty }\Psi _{+}(x) &=&e^{ikx}+R_{L\rightarrow
R}e^{-ikx}\;, \\
\lim_{x\rightarrow +\infty }\Psi _{+}(x) &=&T_{L\rightarrow R}e^{ikx}\;,
\end{eqnarray*}%
we immediately have $c_{+}=1$, $d_{+}=0$ and the transmission and reflection
coefficients turn out to be, respectively 
\begin{eqnarray}
T_{L\rightarrow R} &=&1-i\omega \tilde{g}(k-\alpha )\tilde{h}(k+\beta
)D_{+}(\alpha ,\beta ,k)\,,  \label{TR_LR} \\
R_{L\rightarrow R} &=&-i\omega \tilde{g}(k+\alpha )\tilde{h}(k+\beta
)D_{+}(\alpha ,\beta ,k)\,.  \nonumber
\end{eqnarray}

It is worthwhile to point out that the above expressions break unitarity,
i.e. $\mid T_{L\rightarrow R}\mid^2 +\mid R_{L\rightarrow R}\mid^2 \neq 1$,
because probability flux is not conserved in general.

We come now to the second solution, $\Psi_- (x)$, written in the form 
\begin{equation}
\Psi_-(x) = c_- e^{ikx}+d_- e^{-ikx}+\lambda
I_-(\beta,k)\int_{-\infty}^{+\infty}G_- (x-y)g(y)e^{i\alpha y}dy\, .
\label{Psi2}
\end{equation}
The asymptotic behaviour of the Green's function, $G_- (x)$, is now 
\[
\lim_{x\rightarrow \pm \infty}G_- (x-y)=\frac{i}{2k}e^{\mp ik(x-y)}\,, 
\]
so that 
\[
\lim_{x\rightarrow \pm \infty}\Psi_- (x)=c_- e^{ikx}+d_- e^{-ikx}+i\omega
I_- (\beta,k)\tilde{g}(k\pm \alpha)e^{\mp ikx}\,, 
\]
or, using the explicit expression (\ref{Ipm2}) of $I_-$, 
\begin{eqnarray*}
\lim_{ x\rightarrow -\infty}\Psi_- (x) &=& d_- e^{-ikx}+\left\{c_- +i\omega%
\tilde{g}(k-\alpha) \left[c_-\tilde{h}(k+\beta)+d_-\tilde{h}(k-\beta)\right]%
D_-\right\}e^{ikx}\,, \\
\lim_{ x\rightarrow +\infty}\Psi_- (x) &=& c_- e^{ikx}+\left\{d_- +i\omega%
\tilde{g}(k+\alpha) \left[c_-\tilde{h}(k+\beta)+d_-\tilde{h}(k-\beta)\right]%
D_-\right\}e^{-ikx}\,.
\end{eqnarray*}

Since $\Psi _{-}(x)$ and $\Psi _{+}(x)$ are linearly independent, we can
impose that $\Psi _{-}(x)$ is a wave travelling from right to left, 
\begin{eqnarray*}
\lim_{x\rightarrow -\infty }\Psi _{-}(x) &=&T_{R\rightarrow L}e^{-ikx}\;, \\
\lim_{x\rightarrow +\infty }\Psi _{-}(x) &=&e^{-ikx}+R_{R\rightarrow
L}e^{ikx}\;.
\end{eqnarray*}

The initial conditions \ now are 
\begin{eqnarray*}
c_{-}+i\omega \tilde{g}(k-\alpha )(c_{-}\tilde{h}(k+\beta )+d_{-}\tilde{h}%
(k-\beta ))D_{-}(\alpha ,\beta ,k) &=&0\,, \\
d_{-}+i\omega \tilde{g}(k+\alpha )(c_{-}\tilde{h}(k+\beta )+d_{-}\tilde{h}%
(k-\beta )))D_{-}(\alpha ,\beta ,k)) &=&1\,,
\end{eqnarray*}%
where 
\begin{equation}
d_{-}=T_{R\rightarrow L},\qquad c_{-}=R_{R\rightarrow L}\,.  \label{TRRL}
\end{equation}

We then obtain 
\begin{eqnarray}
T_{R\rightarrow L} &=& 1-i\omega\tilde{g}(k+\alpha)\tilde{h}(k-\beta)%
\mathcal{D}_-(\alpha,\beta,k) \,, \\
R_{R\rightarrow L} &=& -i\omega\tilde{g}(k-\alpha)\tilde{h}(k-\beta)\mathcal{%
D}_-(\alpha,\beta,k) \,.  \nonumber  \label{TR_RL}
\end{eqnarray}
where 
\[
\mathcal{D}_-(\alpha,\beta,k) =\frac{1} {1-\lambda N_-+i\omega (\tilde{g}%
(k+\alpha)\tilde{h}(k-\beta) +\tilde{g}(k-\alpha)\tilde{h}(k+\beta))} \,. 
\]

Formulae (\ref{TR_LR}-\ref{TRRL}) show that, in general, for a $\mathcal{PT}$%
-symmetric non-local potential, $T_{R\rightarrow L}\neq T_{L\rightarrow R}$.
In fact, from the quoted formulae, 
\[
T_{R\rightarrow L}-T_{L\rightarrow R}=i\omega \Delta D_{+}(\alpha ,\beta ,k)%
\mathcal{D}_{-}(\alpha ,\beta ,k)\,,
\]%
where 
\begin{eqnarray*}
\Delta  &=&\tilde{g}(k-\alpha )\tilde{h}(k+\beta )-\tilde{g}(k+\alpha )%
\tilde{h}(k-\beta ) \\
&&+\lambda (N_{+}\tilde{g}(k+\alpha )\tilde{h}(k-\beta )-N_{-}\tilde{g}%
(k-\alpha )\tilde{h}(k+\beta )) \\
&&+i\omega \tilde{g}(k-\alpha )\tilde{h}(k+\beta )(\tilde{g}(k+\alpha )%
\tilde{h}(k-\beta )+\tilde{g}(k-\alpha )\tilde{h}(k+\beta ))\,.
\end{eqnarray*}%
Computation of the $N_{\pm }$ integrals yields the general forms 
\begin{eqnarray*}
N_{+}\left( \alpha ,\beta ,k\right)  &=&\frac{-i}{4k}\left[ \tilde{g}%
(k-\alpha )\tilde{h}(k+\beta )+\tilde{g}(k+\alpha )\tilde{h}(k-\beta )\right]
+Q\left( \alpha ,\beta ,k\right) , \\
N_{-}\left( \alpha ,\beta ,k\right)  &=&\frac{i}{4k}\left[ \tilde{g}%
(k-\alpha )\tilde{h}(k+\beta )+\tilde{g}(k+\alpha )\tilde{h}(k-\beta )\right]
+Q\,\left( \alpha ,\beta ,k\right) ,
\end{eqnarray*}%
where the function $Q\left( \alpha ,\beta ,k\right) $ is real.

If we now make the additional assumption that our kernel is symmetric, $%
K(x,y)=K(y,x)$, i.e. $g=h$ and $\alpha =\beta $, we obtain $T_{R\rightarrow
L}=T_{L\rightarrow R}$ . It is worthwhile to stress that imposing the
symmetry of the kernel is equivalent to imposing the intertwining condition 
\textit{\ }$\mathcal{T}K\mathcal{T}^{-1}=K^{\dagger }$, which ensures the
equality of the two transmission coefficients.

Furthermore, when $\alpha =\beta =0$, the symmetric kernel becomes real, and
exhibits both hermiticity and time reversal invariance. One can then show
that $Q\left( \alpha =0,\beta =0,k\right) $ is a real function of $k$ and
that unitarity holds, \textit{i.e.} $\left\vert T\right\vert ^{2}+\left\vert
R\right\vert ^{2}=1$.

In order to obtain a complete solution of the scattering problem, \textit{%
i.e.} the explicit form of the $Q$ function, we now consider the
one-dimensional $\mathcal{PT}$-symmetric version of a Yamaguchi potential%
\cite{Ya54}, with 
\begin{equation}
g\left( x\right) =\exp \left( -\gamma \left\vert x\right\vert \right)
\;,\qquad h\left( y\right) =\exp \left( -\delta \left\vert y\right\vert
\right) \;,\qquad \left( -\infty <x,y<+\infty \right)  \label{Yagh}
\end{equation}%
where $\gamma $ and $\delta $ are positive numbers. The Fourier transforms
of $g$ and $h$ are, respectively%
\begin{equation}
\tilde{g}\left( q\right) =\frac{2\gamma }{q^{2}+\gamma ^{2}}\;,\qquad \tilde{%
h}\left( q\right) =\frac{2\delta }{q^{2}+\delta ^{2}}\;.  \label{g_h_tild}
\end{equation}

In this case, the $N_{\pm }$ integrals can be computed by elementary
methods, without making use of the Parseval-Plancherel relation and of the
convolution theorem: the $Q$ function can be written in the form%
\begin{eqnarray}
Q\left( \alpha ,\beta ,\gamma ,\delta ,k\right)  &=&\;\frac{2\left( \gamma
^{2}-\alpha ^{2}+k^{2}\right) \left( \gamma +\delta \right) -4\alpha \gamma
\left( \alpha +\beta \right) }{\left[ \left( \gamma +\delta \right)
^{2}+\left( \alpha +\beta \right) ^{2}\right] \left[ \left( \gamma
^{2}-\alpha ^{2}+k^{2}\right) ^{2}+4\alpha ^{2}\gamma ^{2}\right] }
\label{Q_Yama} \\
&&+\frac{1}{4\delta k}\cdot \left[ \left( k-\beta \right) \tilde{g}\left(
k+\alpha \right) \tilde{h}\left( k-\beta \right) +\left( k+\beta \right) 
\tilde{g}\left( k-\alpha \right) \tilde{h}\left( k+\beta \right) \right] \;.
\nonumber
\end{eqnarray}

By inserting the expression of the $Q$ function given above into the $N_{\pm
}$ integrals, we obtain the complete analytic expressions of the two
transmission coefficients, $T_{L\rightarrow R}$ and $T_{R\rightarrow L}$, as
well as the two reflection coefficients, $R_{L\rightarrow R}$ and $%
R_{R\rightarrow L}$, respectively.

Let us put $L\rightarrow R=a$ and $R\rightarrow L=b$ for brevity's sake, and
indicate with $\varphi \left( z\right) $ the phase of the complex number $%
z=\left\vert z\right\vert e^{i\varphi }$, where $z$ represents either a
transmission, or a reflection coefficient. Direct inspection of the formulae
allows us to characterize the behaviour of the coefficients when the kernel
is real, hermitian, or $\mathcal{PT}$-symmetric, summarized in table 2.

It is worthwhile to point out that in the real and hermitian cases, when $%
\left\vert T_{a}\right\vert =\left\vert T_{b}\right\vert =\left\vert
T\right\vert $ and $\left\vert R_{a}\right\vert =\left\vert R_{b}\right\vert
=\left\vert R\right\vert $, unitarity is conserved ($\left\vert T\right\vert
^{2}+\left\vert R\right\vert ^{2}=1$), while in the $\mathcal{PT}$-symmetric
cases ($\left\vert R_{a}\right\vert \neq \left\vert R_{b}\right\vert $)
unitarity is broken ($\left\vert T_{i}\right\vert ^{2}+\left\vert
R_{i}\right\vert ^{2}\neq 1$, with $i=a$, or $b$).

In the case of local $\mathcal{PT}$-symmetric potentials, starting from flux
considerations, a left-right asymmetry in unitarity breaking for a wave
entering the interaction region from the absorptive side ($\Im V<0$),
or from the emissive side ($\Im V>0$) was noticed in Ref.\cite{Ah04} :
while one set of transmission and reflection coefficients obeys the
inequality $\left\vert T_{i}\right\vert ^{2}+\left\vert R_{i}\right\vert
^{2}\leq 1$ for all values of the momentum, $k$, the second set can have $%
\left\vert T_{j}\right\vert ^{2}+\left\vert R_{j}\right\vert ^{2}>1$ for
some values of $k$. Changing the sign of the imaginary part of the $\mathcal{%
PT}$-symmetric potential  while leaving the real part unchanged is
equivalent to a parity transformation, which exchanges left with right and,
consequently, the two sets of coefficients with their asymmetric unitarity
breaking: this property is called handedness in Ref.\cite{Ah04}.

A parity transformation \ of the $\mathcal{PT}$-symmetric Yamaguchi
potential (\ref{K_pm}-\ref{Yagh}) could be obtained either by applying the
definition of $\mathcal{P}$ ($x\rightarrow -x,\;y\rightarrow -y$), or,
equivalently , by the reflection $\alpha \rightarrow -\alpha ,\,\beta
\rightarrow -\beta $. Thus, one can expect that the latter transformation
induces some kind of left-right transformation of scattering observables.
Therefore, one can assert for the $\mathcal{PT}$-symmetric Yamaguchi
potential with real $\lambda $, $\alpha $, $\beta $ and positive $\gamma $, $%
\delta $ the validity of the relations%
\begin{equation}
\begin{array}{c}
T_{L\rightarrow R}\left( \alpha ,\beta \right) =T_{R\rightarrow L}\left(
-\alpha ,-\beta \right) \;, \\ 
R_{L\rightarrow R}\left( \alpha ,\beta \right) =R_{R\rightarrow L}\left(
-\alpha ,-\beta \right) \;.%
\end{array}
\label{ParYama}
\end{equation}

In particular, when $\alpha =\beta =0$, one recovers parity invariance of
the potential. In general, under the $\alpha \rightarrow -\alpha ,\,\beta
\rightarrow -\beta $ transformation, the two transmission coefficients
exchange their phases, while the two reflection coefficients exchange their
moduli. The asymmetry in unitarity breaking is no more valid, \textit{a
priori}, for non-local potentials, and for the Yamaguchi potential in
particular. \ In fact, (\textit{i) }the probability flux does not obey the
standard continuity equation\cite{Ao83} and (\textit{ii}) it is not trivial
to identify an absorptive and an emissive side of the interaction region
unambiguously.

Numerical evaluation of the transmission and reflection coefficients for a
Yamaguchi potential barrier ($\lambda >0$) shows that, when $\alpha $ and $%
\beta $ have opposite sign, $\left\vert T_{i}\right\vert ^{2}+\left\vert
R_{i}\right\vert ^{2}$ may be $\leq 1$ in a range of $k$ values, and $\geq 1$
in another range for 
$i=L\rightarrow R$ and $R\rightarrow L$ simultaneously. In this case, we do not
distinguish an absorptive side and an emissive side any more.

 When both $\alpha $
and $\beta $ are positive, we obtain $\left\vert T_{L\rightarrow
R}\right\vert ^{2}+\left\vert R_{L\rightarrow R}\right\vert ^{2}\leq 1$ and $%
\left\vert T_{R\rightarrow L}\right\vert ^{2}+\left\vert R_{R\rightarrow
L}\right\vert ^{2}\geq 1$ for all values of $k$. Left and right are,
obviously, exchanged under $\alpha \rightarrow -\alpha ,\,\beta \rightarrow
-\beta $, as expected from the parity transformation of the potential.  
Relations (\ref{ParYama}) are, of course, rigorously valid for all values of 
$\alpha $ and $\beta $.

\begin{table}[tbp]
\begin{center}
\begin{tabular}{|l|l|l|l|l|}
\hline
$\alpha =\beta =0$ & $\left\vert T_{a}\right\vert =\left\vert
T_{b}\right\vert $ & $\varphi \left( T_{a}\right) =\varphi \left(
T_{b}\right) $ & $\left\vert R_{a}\right\vert =\left\vert R_{b}\right\vert $
& $\varphi \left( R_{a}\right) =\phi \left( R_{b}\right) $ \\ \hline
$\alpha =-\beta ,\gamma =\delta $ & $\left\vert T_{a}\right\vert =\left\vert
T_{b}\right\vert $ & $\varphi \left( T_{a}\right) \neq \varphi \left(
T_{b}\right) $ & $\left\vert R_{a}\right\vert =\left\vert R_{b}\right\vert $
& $\varphi \left( R_{a}\right) =\phi \left( R_{b}\right) $ \\ \hline
$\alpha =\beta \neq 0,\gamma =\delta $ & $\left\vert T_{a}\right\vert
=\left\vert T_{b}\right\vert $ & $\varphi \left( T_{a}\right) =\varphi
\left( T_{b}\right) $ & $\left\vert R_{a}\right\vert \neq \left\vert
R_{b}\right\vert $ & $\varphi \left( R_{a}\right) =\phi \left( R_{b}\right) $
\\ \hline
$\alpha \neq \beta ,\gamma \neq \delta $ & $\left\vert T_{a}\right\vert
=\left\vert T_{b}\right\vert $ & $\varphi \left( T_{a}\right) \neq \varphi
\left( T_{b}\right) $ & $\left\vert R_{a}\right\vert \neq \left\vert
R_{b}\right\vert $ & $\varphi \left( R_{a}\right) =\phi \left( R_{b}\right) $
\\ \hline
\end{tabular}%
\end{center}
\caption{Properties of transmission and reflection coefficients for (1) real, 
(2) hermitian, (3) symmetric-$\mathcal{PT}$-symmetric and (4)-$%
\mathcal{PT}$-symmetric kernels. }
\end{table}

\section{Conclusions}

\label{concl}

An explicit construction of a solvable separable complex potential has been
presented and worked out in detail. A particularly notable difference
between local and non-local $\mathcal{PT}$-symmetric potentials is the
non-equality of the phases of the two transmission coefficients in the
non-local case.

While the Yamaguchi kind of kernel we have constructed has a cusp at the
origin ($x=y=0$), it has the merit of permitting analytic calculations of
scattering observables in one dimension. Thus, it provides a very convenient
solvable model for $\mathcal{PT}$-symmetric non-local interactions. A
non-trivial feature of this model is the violation of \ the handedness in
unitarity breaking. 

As an outlook on future work, we mention some topics not discussed in detail
in the present work : \ ($i$) the study of the Yamaguchi potential well ($%
\lambda <0$) and of the corresponding bound states; ($ii$) the analytic
structure of the transmission coefficients (\ref{TR_LR}), which appear to be
rational functions of $k$.

Particular attention should be paid to the zeros of the denominator of $%
T_{i}\left( k\right) $ connected with bound states (imaginary $k$) and to
the zeros of the numerator. If the latter occur at positive values of $k$,
they may produce exotic anomalies in the $k$-dependence, $i.e.$ the
vanishing of $T_{i}\left( k\right) $ at isolated values of $k$. In our
understanding, the anomalies are connected with the asymptotic vanishing of
the Wronskian of the solutions, $\Psi _{+}$ and $\Psi _{-}$. The physical
interpretation of such a pathological effect deserves further investigation.

\end{document}